# A Silicon Photonic Accelerator for Convolutional Neural Networks with Heterogeneous Quantization

Febin Sunny, Mahdi Nikdast, and Sudeep Pasricha
Department of Electrical and Computer Engineering,
Colorado State University, Fort Collins, CO, USA
{febin.sunny, mahdi.nikdast, sudeep}@colostate.edu

## ABSTRACT

Parameter quantization in convolutional neural networks (CNNs) can help generate efficient models with lower memory footprint and computational complexity. But, homogeneous quantization can result in significant degradation of CNN model accuracy. In contrast, heterogeneous quantization represents a promising approach to realize compact, quantized models with higher inference accuracies. In this paper, we propose *HQNNA*, a CNN accelerator based on non-coherent silicon photonics that can accelerate both homogeneously quantized and heterogeneously quantized CNN models. Our analyses show that *HQNNA* achieves up to 73.8x better energy-per-bit and 159.5x better throughput-energy efficiency than state-of-the-art photonic CNN accelerators.

## 1. INTRODUCTION

Artificial neural networks (ANNs), especially convolutional neural networks (CNNs), have gained popularity as alternatives to classic machine learning (ML) algorithms. CNNs have exhibited success across application domains such as image and video classification, object detection, and even sequence learning. As CNNs continue to be used for solving increasingly complex problems, they have in turn become even more compute and memory intensive. Research into exploring how to reduce the memory footprint of CNN models while retaining their inference accuracy has been an active area of research in recent years. Some examples of such research areas include exploiting sparsity in CNNs [1], where unnecessary model parameters are pruned, and inducing quantization in CNN models [2], [3], where the parameter bitwidths are reduced. Through quantization, both memory usage and energy requirement for CNN inference can be reduced. Increasing CNN model complexity also necessitates that the underlying hardware platform consistently delivers better performance while satisfying strict energy requirements. Therefore, CNN model optimizations, such as sparsity and quantization are being considered for emerging accelerator platform designs [4].

But even with optimization at the hardware and software levels, electronic CNN accelerators are still prone to diminishing energy and throughput efficiencies, due to the slow-down of Dennard scaling. A potential solution to obtain better energy and throughput efficiency for CNN applications is to consider more efficient hardware technologies, such as silicon photonics, for the design of CNN accelerators. Silicon photonics not only enables low-latency and high bandwidth communication [5], [6], but can also be used for low-latency and energy-efficient computations, e.g., matrix-vector multiplication in the photonic domain [7], which has a computation complexity of only *O(1)*. However, there are various challenges when designing an energy-efficient silicon photonic CNN accelerator, including high laser power, high power dissipation at electro-photonic interfaces, and high latencies associated with inevitable photonic device tuning. Moreover, none of the photonic CNN accelerators proposed to date support the execution of heterogeneously quantized CNN models.

In this work, we propose *HQNNA* which is a silicon photonic CNN accelerator designed for optimizing both homogeneous and heterogeneous quantization in CNN models for energy- and throughput-efficient inference acceleration with high accuracy. Our novel contributions in this work include:

- The design of a novel non-coherent silicon photonic accelerator which utilizes wavelength-division multiplexing (WDM) along with time-division multiplexing (TDM) for bit-slicing-based operation for heterogeneously quantized CNN acceleration;
- The design of energy- and throughput-energy efficient modular vector-granularity-aware matrix-vector multiplication;
- A comprehensive comparison with state-of-the-art silicon-photonic-based CNN accelerators.

The rest of this paper is organized as follows: Section 2 discusses related work and our motivation for this work. Section 3 discusses our *HQNNA* architecture. Section 4 describes the experiments and results. Lastly, Section 5 concludes the work.

## 2. BACKGROUND AND MOTIVATION

Silicon-photonic-based ML accelerator architectures represent an emerging paradigm and can be broadly divided into two major categories: coherent and non-coherent. Due to the superior scalability and performance of non-coherent architectures over coherent architectures [8], the architecture we consider in this work is a non-coherent architecture. Non-coherent architectures use multiple wavelengths, and parameters are imprinted onto the wavelength amplitude by using wavelength-selective devices (e.g., microring resonators (MRs)). Several prior works have discussed CNN acceleration using non-coherent photonic principles. In [11], an MR-based CNN accelerator architecture was proposed which utilizes modular vector-dot-product units with optimized MR designs and tuning circuit optimization, for energy and throughput efficiency. The work in [12] utilized microdisks instead of MRs for lower area and power consumption. Another microdisk-based photonic accelerator was proposed in [13] for fully binarized CNNs (single-bit weight and activation parameters). The work in [14] proposed an MR-based partially binarized CNN accelerator. The partially binarized CNNs allowed for increased inference accuracy over fully binarized CNNs.

For achieving improved memory and computational efficiency, conventional quantization approaches use the same bit-width for all the weight and activation parameters across layers (*homogeneous quantization*). *Heterogeneous or mixed precision quantization* allows different layers to have different levels of quantization to achieve lower memory and computational complexity for similar model accuracy. Several efforts have proposed intelligent neural network architecture search strategies for optimizing the quantization levels across layers in a CNN. A differentiable neural architecture search (DNAS) framework was proposed in [2] to explore the search space with gradient-based optimization. The technique presented in [3] is similar to the one in [2], with an optimized loss function, which penalizes a higher weighted average of the bitwidths of the weights across layers.

Given the prominence of quantized models to achieve efficient CNN deployment on resource-limited embedded and IoT platforms, the ability to accelerate heterogeneously quantized models is essential for modern CNN accelerator architectures. The photonic architectures discussed, support fixed parameter resolution and hence are unable to accelerate heterogeneously quantized models altogether or effectively accelerating heterogeneously quantized models. To fully exploit quantization for latency and energy benefits, we propose the *HQNNA* accelerator, which utilizes WDM and TDM, along with bit-slicing to achieve efficient inference performance.

## 3. HQNNA HARDWARE ACCELERATOR
### 3.1. TDM-based Operation and Energy Benefits

Due to large power consumption required for high resolution DACs and the presence of heterodyne signal crosstalk noise [15], most non-coherent photonic architectures opt to support low-resolution parameters in CNNs. For example, the photonic accelerator discussed in [12] is designed for a 4-bit resolution, while those in [13] and [14] target 1-bit resolution. However, without sufficient optimization of the CNN model, the quantized CNN may exhibit poor inference accuracy at low resolutions, as observed in [12]-[14]. The photonic accelerator in [11] proposed various optimizations in terms of tuning and WDM management to achieve a high resolution of 16-bits, which ensures better inference accuracy than [12]-[14]. However, such an architecture is at a disadvantage in terms of energy efficiency when accelerating a heterogeneously quantized model.

To support heterogeneous quantization and obtain the energy and power benefits it offers, we propose a novel bit-slicing and TDM-based approach in *HQNNA*. Moreover, *HQNNA* makes use of WDM-based operations along with TDM and bit-slicing to aggressively reduce power and energy consumption. Our approach distributes bit-slices across time steps onto the matrix-vector multiplication unit (MVU) to perform the multiplication and accumulation operations photonically, and then makes use of digital shift and adder circuits to obtain the correct output from the MVU operation. The number of time slices required to complete an operation depends on the bit-slice size ($b$) and the parameter size ($p$). An overview of our operation, making use of a simple example, is shown in Fig. 1, involving multiplication of two 2-element ($p$ = 8-bit) vectors: $A$ = [0×31, 0×0D] and $B$ = [0×34, 0×14].

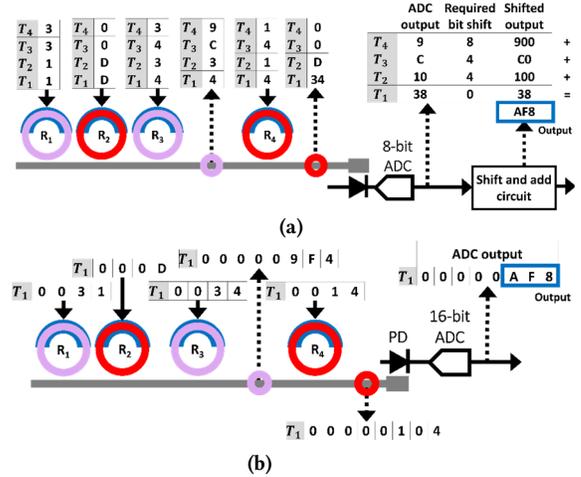

Figure 1: (a) TDM-based operation for a vector-dot-product operation between two 2-element vectors in our proposed *HQNNA* architecture; (b) the same vector-dot-product operation performed with accelerator in [11].

Considering $b$ = 4-bit, *HQNNA* requires four time steps to finish this operation. At *T1*, the least significant nibbles of the elements in A and B are introduced into the multiplication unit. Elements, which must interact with each other during the dot-product operation, are assigned the same wavelength ($\lambda$). The interaction between the nibbles generates intermediate products at each time step, indicated by the colored circles and corresponding callout tables. The intermediate sums (generated using photodetectors) are converted to digital signals using an analog-to-digital converter (ADC), shifted appropriately, and are stored in a local buffer. At *T2*, the second set of nibbles from the *B*-elements are imprinted, while *A* data remains unchanged. After *T2*, all the data from *B* have been introduced to the least significant nibbles of *A* and corresponding sums are obtained. Thus, at *T3*, the second nibble of *A*-elements can be introduced, and *B* needs to be fed again in *T3* and *T4*. [11] will accomplish the same operation in a single time-step (Fig. 1(b)).

The energy consumption for *HQNNA* (Fig. 1(a)) is ~6 $mJ$ while the architecture in [11] (Fig. 1(b)) consumes ~240 $mJ$ for this operation. These energy calculations use the parameters in Table 2, which is discussed later. Even for $p$ of 16-bit (not shown), *HQNNA*, at $b$ of 4-bit, will only have an approximate energy consumption of 24 $mJ$ over 16 time steps, while [11] will still consume 240 $mJ$ (detailed results in Section 4 ).

### 3.2. Tuning Circuits

The thermo-optic (TO) tuning approach is widely used for FPV correction in MR-based systems, and non-coherent photonic accelerator architectures use them for imprinting CNN parameters.

However, the operation of TO tuning circuits can affect the fidelity of operation of neighboring MRs in the form of thermal crosstalk [16]. Therefore, solely relying on microheater-based TO tuning can impair the operation of the non-coherent CNN accelerator. As an alternative, the electro-optic (EO) tuning mechanism operates through carrier injection into the MR body with a PN-junction across the MR. However, the lower tuning range means EO tuning alone is inadequate to address the large variations induced by FPV in MRs but is sufficient for CNN parameter imprinting onto the resonant wavelength. To overcome FPVs and for accurate parameter imprinting required for photonic multiplication, *HQNNA* make use of a hybrid tuning circuit which combines EO and TO tuning. The hybrid tuning approach considers the advantages each tuning mechanism offers while covering for their disadvantages. To address the thermal noise generation from TO tuning, we adapt a method called thermal Eigenmode decomposition (TED), which was first proposed in [16]. TED also comes with the added advantage of significantly reducing TO power consumption and frequency of TO operation.

### 3.3. MVU design

To accelerate ANNs in general, the most time-consuming operation, matrix-vector multiplication, must be accelerated. Inference acceleration in particular deals with fixed weight matrices and input-dependent activations. For CNNs, two main types of layers have to be considered: convolution (CONV) layers and fully connected (FC) layers. CONV layers perform convolution operations between smaller weight matrices or kernels and input feature maps (activations), to generate output feature maps for the next layer. On the other hand, FC layers perform matrix-vector multiplication operations between significantly larger weight matrices and activation vectors. The basic compute unit in our architecture, to support both CONV and FC layer operations, is an MVU. The MVU accepts a WDM signal through an input waveguide, which is imprinted with the vector parameters using an MR bank. For imprinting the parameters, we make use of DAC-based EO tuning in the hybrid tuning circuit. The tuned signal from the MR bank is distributed across the matrix rows, again distributed across waveguides, using a splitter-based photonic multiplexer.

For FC layers, the matrix is comprised of bit-slices of individual weight values, which need to change with time steps so that the vector-matrix multiplication operation can happen in its entirety. For CONV layers, the architecture performs a vector-dot-product operation [11], so the MVU can be used to represent all the bit-slices of one of the vectors simultaneously, across waveguides, to reduce the number of time slices needed for vector-dot-product operations. In both cases, the results per time slice need to be shifted and added. For FC layer operation, this can be done after the summation operation as each waveguide generates partial sums for separate elements. For the CONV layer, the entire MVU generates a single convolution output. In FC layers, the shift and accumulate operation is done electronically, but for CONV layers, it can be done photonically. For photonic shifting, we make use of gain-tuning signal ($\sigma$) fed Semiconductor Optical Amplifiers (SOAs) along with addition via Kirchhoff's Current Law (KCL) from the photodiode outputs.

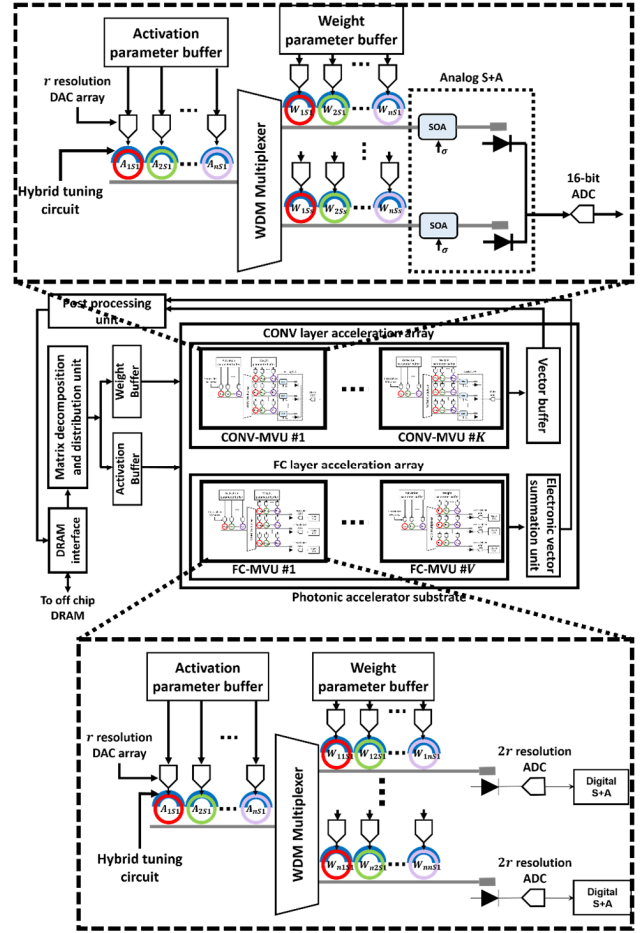

**Figure 2:** Architectural overview of HQNNA with the internal architecture of CONV-MVU and FC-MVU highlighted.

### 3.4. HQNNA Architecture

*HQNNA* architecture, as shown in Fig. 2, is composed of an array of MVUs, with input data routed through an electronic control unit. The MVU array is reused for CONV and FC layer activation. The vectors and matrices are mapped across the MVU array and the resulting partial sum vectors are summed digitally to obtain the sum vectors. For FC layers, each MVU considers an activation vector of size $v$ and a $v \times v$ weight matrix simultaneously. Larger vectors and matrices with different dimensions are split up across different FC-MVUs to obtain the vector to be passed to the next FC layer. The weight parameters must be fed across *ceil(p/b)* time-steps and a single activation vector slice has to operate on all the weight slices. This process needs to be repeated *ceil(p/b)* times to obtain the final output vector. Thus, an output vector of $v$ size is generated every *(ceil(p/b))$^2$* time-steps. For CONV layer acceleration, the kernel, unfurled to a vector of size $k$, and its different bit-slices can be presented simultaneously to a $k$-element, activation vector slice. The activation vector slices, in turn, must be presented to the kernel across *ceil(p/b)* time-steps to obtain a single output vector element. The value of $k$ is decided by the kernel sizes present in the CNN models and may be further decomposed across MVUs as dictated by laser power consumption constraints. As the value of $k$ increases, the MR count, the waveguide length, and hence the laser power needs to be increased, the relation among which can be modeled using:

$$P_{laser} - S_{detector} \geq P_{photoloss} + 10 \times \log_{10} N_\lambda. \quad (1)$$

Table 1: The best models found through heterogeneous quantization techniques considered, compared to the quantized versions from other photonic accelerator works, in terms of inference accuracy and memory footprint

| Model | Number of layers | Parameter count | Quantization type | Weight bitwidths across layers | Activation bitwidths across layers | Inference accuracy | Memory footprint |
|---|---|---|---|---|---|---|---|
| AlexNet | 7 | 38,413,156 | **HQNNA** | [6, 6, 4, 4, 4, 4, 4] | [6, 6, 4, 4, 4, 4, 4] | 76.4% | 169 MB |
| | | | *CrossLight* [11] | 16 | 16 | 79.3% | 650 MB |
| | | | *HolyLight* [12] | 4 | 4 | 76.1% | 162 MB |
| | | | *LightBulb* [13] | 1 | 1 | 56.1% | 41 MB |
| | | | *ROBIN* [14] | 1 | 4 | 62.5% | 48 MB |
| ResNet20 | 20 | 271,786 | **HQNNA** | [2, 2, 2, 2, 2, 2, 2, 2, 2, 2, 2, 2, 2, 2, 4, 2, 2, 2, 2, 4] | [4, 4, 4, 4, 4, 4, 4, 4, 6, 6, 6, 8, 6, 8, 10, 10, 10, 10, 10, 8] | 79.7% | 8.75 MB |
| | | | *CrossLight* [11] | 16 | 16 | 81.9% | 70 MB |
| | | | *HolyLight* [12] | 4 | 4 | 77.6% | 17.5 MB |
| | | | *LightBulb* [13] | 1 | 1 | 56.1% | 4.4 MB |
| | | | *ROBIN* [14] | 1 | 4 | 64.2% | 5.8 MB |
| CNN (SVHN) | 7 | 552,362 | **HQNNA** | [8, 8, 4, 4, 4, 4, 4] | [8, 8, 4, 4, 4, 8, 4] | 87.9% | 34.4 MB |
| | | | *CrossLight* [11] | 16 | 16 | 86.2% | 134 MB |
| | | | *HolyLight* [12] | 4 | 4 | 82.1% | 32.4 MB |
| | | | *LightBulb* [13] | 1 | 1 | 29.4% | 8.4 MB |
| | | | *ROBIN* [14] | 1 | 4 | 49.4% | 9.8 MB |

Here, $P_{laser}$ is the laser power in dBm, $S_{detector}$ is the PD sensitivity in dBm, $N_\lambda$ is the number of wavelengths, and $P_{photoloss}$ is the total optical loss experienced by the signal. We considered optical signal losses due to various factors: waveguide propagation loss (1 dB/cm [11]), splitter loss (0.05 dB [25]), MR through loss (0.02 dB [11]), MR modulation loss (0.72 dB [11]), EO tuning loss (6 dB/cm [9]), and TO tuning power (27.5 mW/FSR [10]). The value of $v$ is more open-ended and needs to be optimized depending on the throughput analysis for FC layers across models. The DAC resolution, and the corresponding power and latency, will be dependent on the $b$ value being used. For CONV layer operations, we consider $K$ Conv-MVUs and for FC layer operation, $V$ FC-MVUs are considered. We analyze the values of these parameters in Section 4.

## 4. EXPERIMENTS AND RESULTS

To evaluate the effectiveness of *HQNNA*, we conducted several simulation-based analyses. For the CNN models, we consider the well-known models AlexNet and ResNet20 for CIFAR 10 dataset classification, along with a custom model for SVHN dataset classification. For power, energy, and latency analysis of silicon photonic CNN accelerators, we developed a Python-based in-house simulator. For analyzing the model accuracy, we used Tensorflow v2.8 along with QKeras [17]. We compare the performance of our architecture in terms of energy-efficiency (energy-per-bit, or EPB), and throughput-energy efficiency (GOPS/EPB) against state-of-the-art photonic CNN accelerators: *CrossLight* [11], *HolyLight* [12], *LightBulb* [13], and *ROBIN* [14]. For obtaining optimal heterogeneous quantization for these models, we explored different algorithms. The best configuration found using [2] was used for AlexNet and ResNet20, and for the SVHN CNN model, an exhaustive quantization search using AutoQKeras was performed (results in Table 1). This quantization exploration among the models, for *HQNNA*, is essentially a search for optimal $p$ value in terms of accuracy and memory footprint. We also simulate the various quantization techniques adapted in the works [11]-[14] that we compare *HQNNA* against (see Table 1). Note how the heterogeneously quantized models (*HQNNA*) have significantly lower memory footprint than the 16-bit quantized models ([11]) while maintaining competitive model accuracy.

Table 2: Parameters considered for architecture analysis

| Devices | Latency | Power |
|---|---|---|
| EO tuning [9] | 20 ns | 4 $\mu W/nm$ |
| TO tuning [10] | 4 $\mu s$ | 27.5 mW/*FSR* |
| VCSEL [18] | 0.07 ns | 1.3 mW |
| Photodetector [19] | 5.8 ps | 2.8 mW |
| SOA [20] | 0.3 ns | 2.2 mW |
| DAC (16-bit) [21] | 0.33 ns | 40 mW |
| ADC (16-bit) [22] | 14 ns | 62 mW |
| DAC (8-bit) [23] | 0.29 ns | 3 mW |
| ADC (8-bit) [24] | 0.82 ns | 3.1 mW |

The power and latency parameters used to model the architectures are shown in Table 2. DACs with lower resolutions (1-bit, 2-bit, 4-bit) are not widely researched, possibly due to niche application spaces. For them, we have assumed the same latency as the design from [24]. We have also scaled DAC power for these lower resolution devices, with respect to resolution ($N$) using the following proportionality [26]:

$$P_{DAC} \propto \left(\frac{2^N}{N} + 1\right). \quad (2)$$

In our first experiment, we optimize the *HQNNA* architecture, in terms of ($v, k, b, V, K$) (see Section 3.4). The best configuration was found in terms of throughput-energy efficiency in terms of GOPS/EPB. The best ($v, k, b, V, K$) was found to be (50, 20, 4, 200, 100) for the CNN models considered. This configuration of *HQNNA* exhibits low maximum-power consumption (57.5 W) due to the lower tuning and DAC power consumption.

Fig. 3 shows the EPB comparison across different architectures.

The lower power consumption of *HQNNA* along with lower latencies of the lower resolution DACs being used enable this architecture to obtain lower EPB values as well. On average, *HQNNA* achieves 73.8×, 52.2×, 12.2×, and 3.59× lower EPBs than *HolyLight*, *LightBulb*, *CrossLight*, and *ROBIN*, respectively, as shown in Fig. 3.

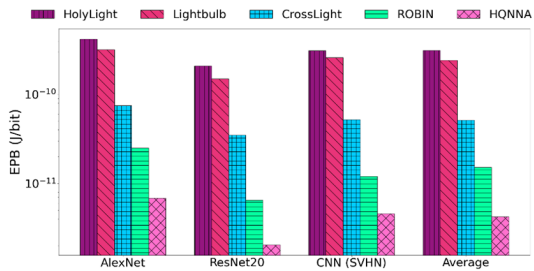

**Figure 3: EPB for CNN models, across photonic accelerators.**

Despite having lower throughput than *CrossLight* and *ROBIN*, due to the significantly lower EPB, *HQNNA* exhibits significantly higher GOPS/EPB. As shown in Fig. 4, our *HQNNA* architecture achieves 159.5×, 103.1×, 28.6×, and 3.37× better GOPS/EPB than *HolyLight*, *LightBulb*, *CrossLight*, and *ROBIN* respectively. These results highlight the energy- and throughput-energy efficient quantized CNN acceleration capabilities of the *HQNNA* accelerator.

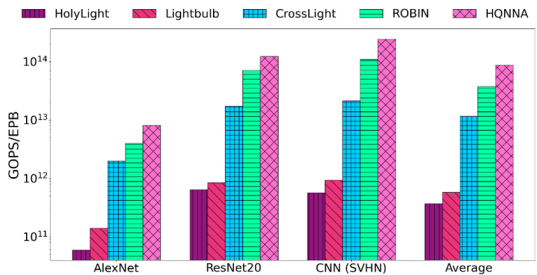

**Figure 4: GOPS/EPB for CNN models, across photonic accelerators.**

## 5. CONCLUSION

In this paper, we presented a novel non-coherent photonic CNN accelerator called *HQNNA*, which uses WDM and TDM simultaneously to efficiently accelerate heterogeneously quantized CNN models. Through identifying optimal quantization profiles for the CNNs and corresponding optimizations for hardware, *HQNNA* succeeded to achieve better performance in terms of energy- and throughput-efficiency: up to 73.8× better energy-per-bit and 159.5× better throughput-energy efficiency than conventional photonic CNN accelerators. Thus, *HQNNA* represents a promising new substrate for energy-efficient quantized CNN model acceleration.

## ACKNOWLEDGEMENTS

This research is supported by grants CCF-1813370 and CCF-2006788 from the National Science Foundation (NSF).